\documentclass[onecolumn,secnumarabic,amssymb, nobibnotes, aps, prd]{revtex4-2}

\usepackage{orcidlink}
\usepackage{subcaption}
\usepackage{graphicx} 
\usepackage{amsmath}
\usepackage{amsthm}
\usepackage{dsfont}
\usepackage[bottom]{footmisc}

\usepackage{chngcntr}
\usepackage{apptools}
\AtAppendix{\counterwithin{lemma}{section}}

\newtheorem{lemma}{Lemma}

\AtAppendix{\counterwithin{theorem}{section}}

\newcommand{\eps}{\varepsilon}
\newcommand{\comm}[2]{\left[#1,#2\right]}

\begin{document}

\eprint{JLAB-THY-24-3986}

\title{Analytic Solutions of the DGLAP Evolution and Theoretical Uncertainties}%

\author{Andrea Simonelli \orcidlink{0000-0003-2607-9004}\,}%
\email[]{andsim@jlab.org}
\affiliation{Department of Physics, Old Dominion University, Norfolk, VA 23529, USA}
\affiliation{
Theory Center, Jefferson Lab, Newport News, Virginia 23606, USA}

\begin{abstract}
    The energy dependence for the singlet sector of Parton Distributions Functions (PDFs) is described by an entangled pair of ordinary linear differential equations. Although there are no exact analytic solutions, it is possible to provide approximated results depending on the assumptions and the methodology adopted. These results differ in their sub-leading, neglected terms and ultimately they are associated with different treatments of the theoretical uncertainties. In this work, a novel analytic approach in Mellin space is presented and a new methodology for obtaining closed and exponentiated analytic solutions is devised.
    Different results for the DGLAP evolution at Next-Lowest-Order are compared, discussing advantages and disadvantages for each solution. The generalizations to higher orders are addressed.
\end{abstract}

\maketitle
\tableofcontents

\section{Introduction\label{sec:intro}}

Parton densities have played a central role in the exploration of the strong force since the introduction of Quantum Chromodynamics (QCD). Over the years, they have undergone extensive scrutiny, resulting in a rich body of literature and numerous achievements in the understanding of hard processes.
Starting from the seminal works by Gribov and Lipatov~\cite{Gribov:1972ri}, Altarelli and Parisi~\cite{Altarelli:1977zs} and Dokshitzer~\cite{Dokshitzer:1977sg} (DGLAP), it became evident that the parton distribution functions (PDFs) are solutions of a set of integro-differential equations, with proper boundary conditions. Specifically, the DGLAP evolution is:
\begin{align}
    \label{eq:DGLAP_eq}
    \frac{\partial}{\partial \log{Q^2}} f_{i/h}(x,Q) = 
    \sum_j \int_x^1 \frac{d z}{z} P_{i/j}(z,a_S(Q)) \,f_{j/h}\left(\frac{z}{x},Q\right) 
\end{align}
where the indices $i,j$ represent any parton species inside the hadron $h$, meaning any available flavor of quarks and anti-quarks at the energy scale $Q$, and the gluon. 
Within this picture, they carry a fraction $x$ of the overall momentum of the hadron. 
The kernels $P_{i/j}$ can be expanded in powers of $\alpha_S$ and the coefficients of such expansion are now completely known up to next-to-next-leading-order (NNLO)~\cite{Moch:2004pa,Vogt:2004mw,Blumlein:2021enk} with some estimates available for the N$^3$LO~\cite{Davies:2016jie,Moch:2017uml,Bonvini:2018xvt,Davies:2022ofz,Moch:2021qrk,Falcioni:2023luc}, which already have been used in very recent extractions~\cite{McGowan:2022nag,Hekhorn:2023gul}. The higher is the energy scale $Q$, the more such expansion is trustful. Realistically, the support of Eq.~\eqref{eq:DGLAP_eq} where perturbative QCD can be confidently applied roughly corresponds to values of $Q$ greater than $\approx 1$ GeV. This lower bound $Q_0$ serves as the reference point for determining boundary conditions, typically derived through phenomenological extraction from experimental data. Once the functional form of $f_{i/h}(x,Q_0)$ is known, the behavior of the parton densities at high energies is forecasted by solving Eq.~\eqref{eq:DGLAP_eq}. 
The methods for obtaining these solutions can be broadly classified into two main categories based on whether the DGLAP equations are solved numerically or analytically.
Each approach has its advantages and drawbacks, and the choice between them depends on the specific objectives of the analysis.
Ultimately, they differ in the treatment and the estimate of the theoretical uncertainties. 

Numerical approaches ensure exact satisfaction of the DGLAP equations and are often implemented directly in $x$-space~\cite{Salam:2008qg,Botje:2010ay,Bertone:2013vaa}. Moreover, numerical solutions are by construction path-independent, meaning that the result at scale $Q$ can be equally obtained by evolving from $Q_0$ to any scale $Q'$ and then from $Q'$ to $Q$. 
As a particular case, the perturbative hysteresis~\cite{Bertone:2022sso} associated to numerical solutions is exactly zero.
In contrast, analytic solutions, while providing transparency, only satisfy these requirements up to a certain error due to the necessary approximations. Analytic solutions are typically more challenging to achieve, particularly as the perturbative order increases. They often involve Mellin space transformations~\cite{Vogt:2004ns,Ball:2008by,Ball:2010de,DelDebbio:2007ee,Moffat:2021dji,Candido:2022tld,NNPDF:2024djq}, introducing additional complexity when handling the inverse Mellin transform. Additionally, different approximations may lead to different analytic solutions of the same equation, affecting the estimation of theoretical uncertainties. 
For instance, the minimization of the theoretical errors could be preferred over a better control of the size of the uncertainties. Alternatively, a simpler analytic result might be more suitable for certain application, even if it comes with bigger errors.
The complexity of analytic methods is compounded by the matrix nature of Eqs.~\eqref{eq:DGLAP_eq}. A suitable change of basis allows to disentangle the evolution for $2 \,n_f - 1$ operators, which build up the non-singlet sector, but this operation still leaves a matrix equation for the singlet sector, constituted by the remaining $2$ operators. The solution for this singlet doublet is challenging beyond the leading order (LO), due to the non-commutativity of the splitting kernels.

Before delving into the methodologies for addressing these challenges, it is essential to clarify the terminology used in this work. Analytic results can be categorized based on their ability to solve the evolution equations, resulting in either \emph{exact} or \emph{approximated} solutions. Exact solutions, by definition, do not violate the equations they solve and have zero perturbative hysteresis, representing the highest level of ambition in solving the evolution equations. Unfortunately, while the non-singlet sector can be successfully solved exactly at any order, exact solutions for the singlet sector can only be obtained at LO, due to its matrix nature. 
Another distinguishing factor is the closure of the solution: if a finite number of operators fully specify the solution, it is considered \emph{closed}; otherwise, if an infinite number of operators are involved in an iterative algorithm, it is termed as \emph{iterated}. Furthermore, solutions can be characterized by their functional form, with exact solutions necessarily being exponentiated.

Beyond LO, the analytic evolution of the singlet sector typically relies on approximated results, commonly obtained through the popular $U$-matrix approach~\cite{Buras:1979yt,Furmanski:1981cw,Vogt:2004ns}. While this method offers simplicity, it compromises the mathematical structure of the result and may not adequately minimize theoretical uncertainties when expressed in a closed form, especially at low energies $Q \approx Q_0$. In this paper, I introduce a novel approach for addressing the singlet sector, which not only enhances mathematical elegance but also improves theoretical precision. Specifically, despite the conventional notion that the non-commutativity of splitting kernels prevents closed exponential solutions beyond NLO~\cite{Blumlein:1997em}, I demonstrate how this can be achieved by leveraging matrix representations in two dimensions. This advancement holds significant implications for current and future parton density phenomenology, particularly in the era of precision physics at the Large Hadron Collider (LHC) and the forthcoming Electron-Ion Collider (EIC) experiments~\cite{Amoroso:2022eow,AbdulKhalek:2021gbh}.

\section{Singlet Evolution \label{sec:singlet}}

It is long-time known that the Eqs.~\eqref{eq:DGLAP_eq} are simpler in Mellin space, where $x$-convolutions are mapped to products. 
The DGLAP equation for the singlet sector becomes:
\begin{align}
    \label{eq:singlet_DGLAP}
    &\frac{\partial}{\partial \log{Q^2}} 
    \begin{pmatrix}
        \Sigma(N,Q) \\
        g(N,Q)
    \end{pmatrix}
    =
    \mathbf{P}\left(N,a_S(Q)\right)
    \begin{pmatrix}
        \Sigma(N,Q) \\
        g(N,Q)
    \end{pmatrix}
\end{align}
where $\Sigma = \sum_i (q_i + \overline{q}_i)$ and $g$ are associated with the flavor-singlet quark distribution and the gluon distribution, respectively.
The doublet at scale $Q$ can be expressed in terms of the doublet at scale $Q_0$ by introducing the Evolution Operator $\mathbf{E}$:
\begin{align}
    \label{singlet_EO}
    &\begin{pmatrix}
        \Sigma(N,Q) \\
        g(N,Q)
    \end{pmatrix}
    =
    \mathbf{E}\left(N; Q_0, Q\right)
    \,
    \begin{pmatrix}
        \Sigma(N,Q_0) \\
        g(N,Q_0)
    \end{pmatrix}.
\end{align}
Thus, Eq.~\eqref{eq:singlet_DGLAP} can be regarded as the evolution equation for the Evolution Operator itself:
\begin{align}
    \label{eq:singlet_EOevo}
    &\frac{\partial}{\partial \log{Q^2}} 
    \mathbf{E}\left(N; Q_0, Q\right)
    =
    \mathbf{P}\left(N,a_S(Q)\right)
    \mathbf{E}\left(N; Q_0, Q\right),
\end{align}
given the initial condition $\mathbf{E}\left(N; Q_0, Q_0\right) = \mathbf{1}$, the $2\times2$ identity matrix. 

\subsection{Evolution in 2 Dimensions\label{ssec:evo2dim}}

The linear ordinary differential equation of Eq.~\eqref{eq:singlet_EOevo} belongs to a much wider family of initial value problems, which are extremely common in many areas of Physics, not least because its structure encodes the time evolution for a quantum system with Hamiltonian $H$:
\begin{align}
    \label{eq:time_evo}
    &\frac{\partial \widehat{U}(t)}{\partial t}
    =
    \widehat{H}(t) \widehat{U}(t)\,,
    \qquad \widehat{U}(0) = 1.
\end{align}
Here, $\widehat{U}$ and $\widehat{H}$ are generic operators acting on some Hilbert space. The general formal solution is given in terms of the Dyson time-ordered exponential~\cite{Dyson:1949bp}: 
\begin{align}
    \label{eq:time_evo_formalsol}
    &\widehat{U}(t) = \mathcal{T}\text{exp}\left(
    \int_0^t d\tau \widehat{H}(\tau)
    \right) = 
    1 + \int_0^t d\tau \widehat{H}(\tau)
    + \frac{1}{2} \int_0^t d\tau_1 \int_0^{\tau_1} d\tau_2 \widehat{H}(\tau_1) \widehat{H}(\tau_2) + \dots
\end{align}
that reduces to standard exponentiation if the Hamiltonian computed at different times commute with each other.
This solution has been the foundation of modern Quantum Field theory, since it is well suited for the development of perturbative expansions. However, it becomes cryptic and difficult to be used without explicitly expanding the exponential operator. 
An alternative to Eq.~\eqref{eq:time_evo_formalsol} is given by the Magnus expansion~\cite{Magnus:1954zz}:
\begin{align}
    \label{eq:Magnus_exp}
    \widehat{U}(t) = e^{\widehat{\Omega}(t)},
    \quad\text{with }
    \widehat{\Omega}(t) = \sum_{k\geq 1}\widehat{\Omega}_k(t).
\end{align}
The terms $\widehat{\Omega}_k$ involve nested commutators of the Hamiltonian at different times and their complexity increases rather fast. The first two terms are:
\begin{subequations}
\label{eq:Magnus_coeffs}
\begin{align}
    &\widehat{\Omega}_1(t) = \int_0^t d\tau \widehat{H}(\tau),
    \\
    &\widehat{\Omega}_2(t) = \frac{1}{2}\int_0^t d \tau_1 \int_0^{\tau_1} d\tau_2 
    \comm{\widehat{H}(\tau_1)}{\widehat{H}(\tau_2)}.
\end{align}   
\end{subequations}
Terms up to $k=4$ are still manageable, but higher orders become unwieldy.
A comprehensive review of the Magnus expansion and its application can be found in Ref.~\cite{Blanes:2008xlr}, from which part of the language and the nomenclature of this Section has been borrowed.

Although the level of complexity of Eq.~\eqref{eq:time_evo_formalsol} and Eq.~\eqref{eq:Magnus_exp} is rather similar, the latter has the advantage of preserving explicitly the exponential nature of the solution of the differential equation. However, it still require to face the non-commutativity of the Hamiltonian at different times. 
The problem simplifies if the Hamiltonian can be split as $\widehat{H}_0(t)  + \eps \widehat{H}_1(t)$, with $\eps \ll 1$ a small perturbation parameter, which is a situation common to many physical systems. if $\widehat{H}_0$ is diagonalizable, then the following preliminary linear transformation is applied to the operator $\widehat{U}$:
\begin{align}
    \label{eq:prelim_lineartr}
    &\widehat{U}(t) = \widehat{G}(t) \widehat{U}_{\text{\footnotesize int}}(t) \widehat{G}(0)^{-1},
    \quad\text{with }
    \widehat{G}(t) = \text{exp}\left(\int_0^t d\tau \widehat{H}_0(\tau)\right)
\end{align}
and the new interacting operator $\widehat{U}_{\text{\footnotesize int}}$ satisfies the following evolution equation:
\begin{align}
    \label{eq:UG_evo}
    \frac{\partial \widehat{U}_{\text{\footnotesize int}}(t)}{\partial t}
    = \eps \, \widehat{H}_{\text{\footnotesize int}}(t) \widehat{U}_{\text{\footnotesize int}}(t),
    \quad\text{ where }
    \widehat{H}_{\text{\footnotesize int}}(t) =  
    \text{exp}\left(-\widehat{S}_0(t)\right) \widehat{H}_1(t) 
    \,
    \text{exp}\left(\widehat{S}_0(t)\right)
\end{align}
and we have defined the operator $\widehat{S}_0(t) = \int_0^t d\tau \widehat{H}_0(\tau)$.
Introducing the analogous operator $\widehat{S}_1(t) = \int_0^t d\tau \widehat{H}_1(\tau)$, the final solution would be formally given by:
\begin{align}
    \label{eq:U_formalsol}
    &\widehat{U}(t) = 
    \text{exp}\left(\widehat{S}_0(t)\right)
    \text{exp}\left(\eps \, \widehat{S}_1(t)\right)
    \mathcal{T}\text{exp}\left(\eps \, 
    \int_0^t 
    d\tau 
    e^{-\eps\,\widehat{S}_1(\tau)}
    \left(\widehat{H}_{\text{\footnotesize int}}(\tau) - \widehat{H}_1(\tau)\right)
    e^{\eps\,\widehat{S}_1(\tau)}\right).
\end{align}
Despite these manipulations, in general the result above is still difficult to deal with, and simpler analytic forms are hard to be found, except for very few special cases.

A huge improvement in transparency can be achieved if the operators belongs to some bi-dimensional representation. In this case, all the operators are $2\times2$ matrices (denoted in bold case) and we can take advantage of the results collected in Appendix~\ref{app:mat2x2}. 
The interacting Hamiltonian becomes:
\begin{align}
    \label{eq:HG_2dim}
    \mathbf{H}_{\text{\footnotesize int}}(t) = 
    \mathbf{H}_1(t) - 
    \frac{\sinh{\left(\Delta_{S_0}(t)\right)}}{\Delta_{S_0}(t)}
    \comm{\mathbf{S}_0(t)}{\mathbf{H}_1(t)}
    +
    \frac{\cosh{\left(\Delta_{S_0}(t)\right)-1}}{\Delta^2_{S_0}(t)}
    \comm{\mathbf{S}_0(t)}{\comm{\mathbf{S}_0(t)}{\mathbf{H}_1(t)}}
\end{align}
where $\Delta_{S_0}$ is the difference of the eigenvalues of the matrix $\mathbf{S}_0$. Notice that if $|\Delta_{S_0}| \gg 1$, then the last two terms in the expression above might be large and comparable to $\mathbf{H}_0$. 
The $2$-dimensional version of Eq.~\eqref{eq:U_formalsol} is:
\begin{align}
    \label{eq:U_2dimsol}
    &\mathbf{U}(t) = 
    \text{exp}\left(\mathbf{S}_0(t)\right)
    \text{exp}\left(\eps \, \mathbf{S}_1(t)\right)
    \mathcal{T}\text{exp}\Bigg(\eps \, 
    \int_0^t 
    d\tau 
    \Big(
    \mathbf{H}_{\text{\footnotesize int}}(\tau) - \mathbf{H}_1(\tau)
    -
    \frac{\sinh{\left(\eps \Delta_{S_1}(\tau)\right)}}{\Delta_{S_1}(\tau)}
    \comm{\mathbf{S}_1(\tau)}{\mathbf{H}_{\text{\footnotesize int}}(\tau) - \mathbf{H}_1(\tau)}
    +
    \notag \\
    &\quad+
    \frac{\cosh{\left(\eps \Delta_{S_1}(\tau)\right)}-1}{\Delta_{S_1}(\tau)}
    \comm{\mathbf{S}_1(\tau)}{\comm{\mathbf{S}_1(\tau)}{\mathbf{H}_{\text{\footnotesize int}}(\tau) - \mathbf{H}_1(\tau)}}
    \Big)
    \Bigg)
\end{align}
where $\Delta_{S_1}$ is the difference of the eigenvalues of the matrix $\mathbf{S}_1$. The time-ordered exponential above can now be re-casted by using the Magnus expansion as in Eq.~\eqref{eq:Magnus_exp}. 
If $\eps$ really is a small parameter, we can approximate the result as:
\begin{align}
    \label{eq:U_2dimappr}
    &\mathbf{U}^{\text{\footnotesize appr}}(t) = 
    \text{exp}\left(\mathbf{S}_0(t)\right)
    \,
    \text{exp}\left(\eps \, \mathbf{S}_1(t)\right)
    \,
    \text{exp}\left(\eps \, \mathbf{T}_1(t)\right)
    \,
    \text{exp}\left(\eps \, \mathbf{T}_2(t)\right)
\end{align}
where we have introduced the matrices:
\begin{subequations}
    \label{eq:T_matrices}
    \begin{align}
    &\mathbf{T}_1(t) =
    -\int_0^t d \tau  
    \frac{\sinh{\left(\Delta_{S_0}(\tau)\right)}}{\Delta_{S_0}(\tau)}
    \comm{\mathbf{S}_0(\tau)}{\mathbf{H}_1(\tau)} ,
    \\
    &\mathbf{T}_2(t) =
    \int_0^t d \tau  
    \frac{\cosh{\left(\Delta_{S_0}(\tau)\right)-1}}{\Delta_{S_0}(\tau)}
    \comm{\mathbf{S}_0(\tau)}{\comm{\mathbf{S}_0(\tau)}{\mathbf{H}_1(\tau)}}.
    \end{align}
\end{subequations}
The expression in Eq.~\eqref{eq:U_2dimappr} is incredibly powerful, as it states that, whenever the hypothesis on which it is based are satisfied, the first order correction to the solution of the differential equation in Eq.~\eqref{eq:time_evo} induced by the perturbation $\mathbf{H}_1$ in two dimensions is completely determined by solely four matrices.

\bigskip

The strategy sketched above can be generalized to include higher order corrections. 
The key is isolating exponentials of single matrices applying consecutively the Zassenhaus formula and the Magnus expansion.
The number of operators required to completely determine the solution of the initial value problem in Eq.~\eqref{eq:time_evo} increases going to higher orders, but it remains finite and under control.
For instance, a further correction $\eps^2 \mathbf{H}_2(t)$ to the initial Hamiltonian would require a solution approximated up to order $\eps^2$.
\begin{align}
    \label{eq:U_sol_2dim_eps2}
    \mathbf{U}^{\text{\footnotesize appr}}(t) = 
    \text{exp}\left(\mathbf{S}_0(t)\right)
    \text{exp}\left(\eps \,\mathbf{S}_1(t)\right)
    \prod_{i=1}^2
    \text{exp}\left(\eps \, \mathbf{T}_i(t)\right)
    \text{exp}\left(\eps^2 \,\mathbf{S}_2(t)\right)
    \prod_{i=1}^2
    \text{exp}\left(\eps^2 \, \mathbf{Q}_i(t)\right)
    \prod_{i}
    \text{exp}\left(\frac{1}{2}\eps^2 \, \mathbf{W}_i(t)\right)
\end{align}
where, in addition to the four matrices already introduced at order $\eps$, we have defined the following further nine operators:
\begin{subequations}
    \begin{align}
    &\mathbf{S}_2(t) = \int_0^t d\tau \mathbf{H}_2(t),
    \\
    &\mathbf{Q}_1(t) =  -\int_0^t d \tau  
    \frac{\sinh{\left(\Delta_{S_0}(\tau)\right)}}{\Delta_{S_0}(\tau)}
    \comm{\mathbf{S}_0(\tau)}{\mathbf{H}_2(\tau)} ,
    \\
    &\mathbf{Q}_2(t) =
    \int_0^t d \tau  
    \frac{\cosh{\left(\Delta_{S_0}(\tau)\right)-1}}{\Delta^2_{S_0}(\tau)}
    \comm{\mathbf{S}_0(\tau)}{\comm{\mathbf{S}_0(\tau)}{\mathbf{H}_2(\tau)}},
    \\
    &\mathbf{W}_1(t) = 
    \int_0^t d\tau_1 \int_0^{\tau_1} d\tau_2 \comm{\mathbf{H}_1(\tau_1)}{\mathbf{H}_1(\tau_2)},
    \\
    &\mathbf{W}'_1(t) = 
    -\left(\int_0^t d\tau_1 \int_0^{\tau_1} d\tau_2
    -
    \int_0^t d\tau_2 \int_0^{\tau_2} d\tau_1
    \right)
    \frac{\sinh{\Delta_{S_0}(\tau_1)}}{\Delta_{S_0}(\tau_1)}
    \comm{\mathbf{H}_1(\tau_1)}{\comm{\mathbf{S}_0(\tau_2)}{\mathbf{H}_1(\tau_2)}},
    \\
    &\mathbf{W}''_1(t) = 
    \left(\int_0^t d\tau_1 \int_0^{\tau_1} d\tau_2
    -
    \int_0^t d\tau_2 \int_0^{\tau_2} d\tau_1
    \right)
    \frac{\cosh{\Delta_{S_0}(\tau_1)}-1}{\Delta^2_{S_0}(\tau_1)}
    \comm{\mathbf{H}_1(\tau_1)}{\comm{\mathbf{S}_0(\tau_2)}{\comm{\mathbf{S}_0(\tau_2)}{\mathbf{H}_1(\tau_2)}}},
    \\
    &\mathbf{W}_2(t) = 
    \int_0^t d\tau_1 
    \int_0^{\tau_1} d\tau_2 
    \frac{\sinh{\Delta_{S_0}(\tau_1)}}{\Delta_{S_0}(\tau_1)}
    \frac{\sinh{\Delta_{S_0}(\tau_2)}}{\Delta_{S_0}(\tau_2)}
    \comm{\comm{\mathbf{S}_0(\tau_1)}{\mathbf{H}_1(\tau_1)}}{\comm{\mathbf{S}_0(\tau_2)}{\mathbf{H}_1(\tau_2)}}
    \\
    &\mathbf{W}'_2(t) = 
    -\left(\int_0^t d\tau_1 
    \int_0^{\tau_1} d\tau_2
    -
    \int_0^t d\tau_2 
    \int_0^{\tau_2} d\tau_1
    \right)
    \frac{\sinh{\Delta_{S_0}(\tau_1)}}{\Delta_{S_0}(\tau_1)}
    \frac{\cosh{\Delta_{S_0}(\tau_2)}-1}{\Delta_{S_0}(\tau_2)}
    \notag \\
    &\qquad\times
    \comm{\comm{\mathbf{S}_0(\tau_1)}{\mathbf{H}_1(\tau_1)}}{\comm{\mathbf{S}_0(\tau_2)}{\comm{\mathbf{S}_0(\tau_2)}{\mathbf{H}_1(\tau_2)}}}
    \\
    &\mathbf{W}_3(t) = 
    \int_0^t d\tau_1 
    \int_0^{\tau_1} d\tau_2
    \frac{\cosh{\Delta_{S_0}(\tau_1)}-1}{\Delta_{S_0}(\tau_1)}
    \frac{\cosh{\Delta_{S_0}(\tau_2)}-1}{\Delta_{S_0}(\tau_2)}
    \comm{\comm{\mathbf{S}_0(\tau_1)}{\comm{\mathbf{S}_0(\tau_1)}{\mathbf{H}_1(\tau_1)}}}{\comm{\mathbf{S}_0(\tau_2)}{\comm{\mathbf{S}_0(\tau_2)}{\mathbf{H}_1(\tau_2)}}}
    \end{align}
\end{subequations}
The functions $\mathbf{W}_i$ are the terms associated with the first non-trivial contribution of the Magnus expansion.

\subsection{Closed Exponentiated Solution at Next-Lowest-Order \label{ssec:singlet_nlo}}

The formalism outlined above can now easily applied to the solution of the DGLAP evolution at NLO. The starting point is Eq.~\eqref{eq:singlet_EOevo}, with the splitting matrix $\mathbf{P}$ expanded up to order $a_S^2$:
\begin{align}
    \label{eq:singlet_EOevo_nlo}
    \frac{\partial}{\partial \log{Q^2}} \mathbf{E}\left(N;Q\right)
    =
    \left(a_S(Q) \mathbf{P}_0(N) + a_S^2(Q) \mathbf{P}_1(N)\right)
    \mathbf{E}\left(N;Q\right).
\end{align}
Expressions are simpler if we treat $a_S(Q)$ as the independent variable. Hence, we map the equation above as:
\begin{align}
    \label{eq:singlet_EOevo_nloR}
    \frac{\partial}{\partial a_S} \mathbf{E}\left(N;a_S\right)
    =
    \left(
    -\frac{1}{a_S} \mathbf{R}_0(N)
    -\frac{1}{1+a_S b_1} \mathbf{R}_1(N)
    \right)
    \mathbf{E}\left(N;a_S\right).
\end{align}
The original equation is recovered by multiplying each side by the QCD beta function at NLO $\beta(a_S) = - \beta_0 a_S^2 - \beta_1 a_S^3$ and setting $b_1 = \beta_1/\beta_0$ together with the definitions:
\begin{align}
    \label{eq:R_operators}
    &\mathbf{R}_0 = \frac{\mathbf{P}_0}{\beta_0};
    \qquad
    \mathbf{R}_1 = \frac{\mathbf{P}_1}{\beta_0} - b_1 \mathbf{R}_0 .
\end{align}
The “Hamiltonian" associated with this evolution is:
\begin{align}
    \mathbf{H}(a_S) =
    -\frac{1}{a_S} \mathbf{R}_0(N)
    -\frac{1}{1+b_1 a_S} \mathbf{R}_1(N),
\end{align}
and the second term can be considered as a perturbation when $a_S$ is a small parameter, which is always assumed to be satisfied if $Q \geq Q_0$.
It is now quite straightforward to write the solution as in Eq.~\eqref{eq:U_2dimappr}. We simply have to compute the four matrices:
\begin{subequations}
\label{eq:ingredients_NLO}
    \begin{align}
    \label{eq:singlet_NLOsol}
    &\mathbf{S}_0\left(N;a_0,a_S\right) = 
    -\int_{a_0}^{a_S} \frac{d a}{a} \mathbf{R}_0(N)
    =
    h_1\left(a_0,a_S\right)\mathbf{R}_0(N);
    \\
    &\mathbf{S}_1\left(N;a_0,a_S\right) = 
    -\int_{a_0}^{a_S} \frac{d a}{1 + b_1 a} \mathbf{R}_1(N)
    =
    h_2\left(a_0,a_S\right)
    \mathbf{R}_1(N);
    \\
    &\mathbf{T}_1\left(N;a_0,a_S\right)
    =
    \int_{a_0}^{a_S} \frac{d a}{1 + b_1 a} 
    \frac{\sinh{\left(\Delta_0(N)
     h_1(a_0,a) \right)}}{\Delta_0(N)}
    \comm{\mathbf{R}_0(N)}{\mathbf{R}_1(N)}
    =
    \notag \\
    &\quad=
    h_3\left(\Delta_0(N); a_0,a_S\right)
    \comm{\mathbf{R}_0(N)}{\mathbf{R}_1(N)};
    \\
    &\mathbf{T}_2\left(N;a_0,a_S\right)
    =-\int_{a_0}^{a_S} \frac{d a}{1 + b_1 a} 
    \frac{\cosh{\left(\Delta_0(N)
    h_1(a_0,a)  \right)-1}}{\Delta_0^2(N)}
     \comm{\mathbf{R}_0(N)}{ \comm{\mathbf{R}_0(N)}{\mathbf{R}_1(N)}}
     =
     \notag \\
     &\quad=
     h_4\left(\Delta_0(N); a_0,a_S\right)
    \comm{\mathbf{R}_0(N)}{ \comm{\mathbf{R}_0(N)}{\mathbf{R}_1(N)}},
\end{align}
\end{subequations}
where $a_0 = a_S(Q_0)$, $\Delta_0$ is the difference of the eigenvalues of $\mathbf{R}_0$, and the functions $h_i$ are given by:
\begin{subequations}
    \label{eq:ingredients_NLO_funs}
    \begin{align}
    &h_1\left(a_0,a_S\right) = -\log{\left(\frac{a_S}{a_0}\right)};
    \\
    &h_2\left(a_0,a_S\right) = -\frac{1}{b_1}\log{\left(\frac{1 + b_1 a_S}{1 + b_1 a_0}\right)};
    \\
    &h_3\left(\Delta; a_0,a_S\right)=
     -\frac{1}{2 \Delta}\left(a_S F_-(\Delta;a_0,a_S)-
    a_0 F_-(\Delta;a_0,a_0)\right);
    \\
    &h_4\left(\Delta; a_0,a_S\right)=
    -\frac{1}{2\Delta^2}
     \left(a_S F_+(\Delta;a_0,a_S)-
    a_0 F_+(\Delta;a_0,a_0) 
    +2 h_2(a_0,a_S)\right),
    \end{align}
\end{subequations}
together with:
\begin{align}
    \label{eq:Ffun_def}
    &F(\Delta;a_0,a_S) = 
    \left(\frac{a_S}{a_0}\right)^{\Delta}
    \frac{1}{1+\Delta}{}_{2}F_{1}\left(1,1+\Delta;2+\Delta,-b_1 a\right); \\
    &F_+(\Delta;a_0,a_S) = F(\Delta;a_0,a_S)+F(-\Delta;a_0,a_S);
    \\
    &F_-(\Delta;a_0,a_S) = F(\Delta;a_0,a_S)-F(-\Delta;a_0,a_S),
\end{align}
with ${}_{2}F_{1}$ being the Gauss Hypergeometric function. 
Finally, the solution for the singlet sector evolution operator obtained by following the recipe of Section~\ref{ssec:evo2dim} reads as:
\begin{align}
    \label{eq:singlet_EO_solNLO}
    &\mathbf{E}^{\text{\footnotesize NLO}}\left(N; a_0,a_S\right)
    =
    \text{exp}\left(h_1(a_0,a_S) \mathbf{R}_0(N)\right)
    \text{exp}\left(h_2(a_0,a_S) \mathbf{R}_1(N)\right)
    \notag \\
    &\quad\times
    \text{exp}\left(h_3(\Delta_0(N); a_0,a_S)  \comm{\mathbf{R}_0(N)}{\mathbf{R}_1(N)}\right)
    \text{exp}\left(h_4(\Delta_0(N); a_0,a_S) \comm{\mathbf{R}_0(N)}{ \comm{\mathbf{R}_0(N)}{\mathbf{R}_1(N)}}\right).
\end{align}
In the following, we will refer to this solution as the Analytic Solution at NLO. 
Notice that its practical implementations only require to diagonalize three matrices, namely $\mathbf{R}_0$, $\mathbf{R}_1$ and their commutator. In fact, the $2$-commutator appearing in the last factor is anti-diagonalized by the same matrix that diagonalizes $\comm{\mathbf{R}_0}{\mathbf{R}_1}$.
This result is also consistent with the exact analytic solution in the Non-singlet sector:
\begin{align}
    \label{eq:Nonsinglet_EO_solNLO}
    &E^{\text{\footnotesize
    NLO}}_{\text{\footnotesize NS}}\left(N; a_0,a_S\right)
    =
    \text{exp}\left(
    h_1(a_0,a_S) R^{\footnotesize NS}_0(N)
    +
    h_2(a_0,a_S) R^{\footnotesize NS}_1(N)
    \right),
\end{align}
which is the regular functions version of Eq.~\eqref{eq:singlet_EO_solNLO}.
Notice that the inclusion of QED corrections (even at LO) can be pursued by adopting the same techniques used to get the result of Eq.~\eqref{eq:singlet_EO_solNLO}.

\bigskip

The generalization to higher orders is addressed within the same formalism adopted for treating the NLO. The Eq.\eqref{eq:singlet_EOevo_nloR} at N$^n$LO reads as:
\begin{align}
    \label{eq:singlet_EOevo_nMloR}
    \frac{\partial}{\partial a_S} \mathbf{E}\left(N;a_S\right)
    =
    \left(
    -\frac{1}{a_S} \mathbf{R}_0(N)
    -\sum_{k=1}^n a_S^{k-1}
    \frac{1 + \sum_{j=1}^{n-k} b_j\,a_S^j}{1 + \sum_{j=1}^{n} b_j\,a_S^j} \mathbf{R}_k(N)
    \right)
    \mathbf{E}\left(N;a_S\right).
\end{align}
where $b_k = \beta_k/\beta_0$ and the operators $\mathbf{R}_k$ beyond NLO are defined as:
\begin{align}
\label{eq:R_operator_general}
    \mathbf{R}_k = \frac{\mathbf{P}_k}{\beta_0} - \sum_{j=1}^k b_j \mathbf{R}_{k-j}
\end{align}
Once the “Hamiltonian" associated to the evolution at $n$th-order has been determined, the operators contributing to the matrix exponentials of the solution are obtained following the strategy outlined in Section~\ref{ssec:evo2dim}. For instance, the NNLO is obtained from the expression in Eqs.~\eqref{eq:U_sol_2dim_eps2} and by computing the associated integrations. 

\subsection{Comparison with the $U$-matrix approach}

A popular alternative route for obtaining a solution in the singlet sector is to adopt the following ansatz for the Evolution Operator~\cite{Buras:1979yt,Furmanski:1981cw,Vogt:2004ns}:
\begin{align}
    \label{eq:Umatrices_EO}
    \mathbf{E}\left(N;a_0,a_S\right)
    =
    \mathbf{U}\left(N;a_S\right) \text{exp}\left(h_1(a_0,a_S) \mathbf{R}_0(N)\right)
    \mathbf{U}^{-1}\left(N;a_0\right).
\end{align}
Despite its simple structure and wide usage, the separation of scales presented by the expression above looks suspicious.
In fact, the operator $\mathbf{U}$ depends on a single scale rather than both $a_S$ and $a_0$, in contrast with the general formal solution which involves operators integrated from $Q_0$ to $Q$, and hence necessarily depending on both scales. 
Nevertheless, this peculiar separation is usually enforced by the following expansion in the strong coupling:
\begin{align}
    \label{eq:U_exp}
    \mathbf{U}\left(N;a_S\right)
    =
    1 + \sum_{k=1}^\infty a_S^k \mathbf{U}_k(N).
\end{align}
The problem is then reduced to finding the matrices $\mathbf{U}_k$. Substituting this expansion in the original evolution equation and equating the coefficients order by order, a chain of commutation relations between the matrices $\mathbf{U}_k$ and the operators $\mathbf{R}_k$ defined in Eq.~\eqref{eq:R_operator_general} appears. This leads to the following recursive definition for the matrices $\mathbf{U}_k$:
\begin{align}
   \label{eq:Uk_def}
    \mathbf{U}_k = -\frac{1}{k}\left(
    \mathbf{e}_- \widetilde{\mathbf{R}}_k \mathbf{e}_-
    +
    \mathbf{e}_+ \widetilde{\mathbf{R}}_k \mathbf{e}_+
    \right)
    +
    \frac{\mathbf{e}_+ \widetilde{\mathbf{R}}_k \mathbf{e}_-}{\Delta_0 - k}
    -
    \frac{\mathbf{e}_- \widetilde{\mathbf{R}}_k \mathbf{e}_+}{\Delta_0 + k} 
\end{align}
where $\Delta_0$ is the difference of the eigenvalues of $\mathbf{R}_0$ as in Section~\ref{ssec:singlet_nlo} and $\mathbf{e}_{\pm}$ are related to the matrix $\mathbf{V}_{\text{\footnotesize LO}}$ that diagonalizes the LO operator $\mathbf{R}_0$:
\begin{align}
    \mathbf{e}_-(N) = \mathbf{V}_{\text{\footnotesize LO}}(N) \begin{pmatrix}
        1 && 0 \\
        0 && 0
    \end{pmatrix}
    \mathbf{V}_{\text{\footnotesize LO}}^{-1}(N)\,;
    \qquad
    \mathbf{e}_+(N) = \mathbf{V}_{\text{\footnotesize LO}}(N) \begin{pmatrix}
        0 && 0 \\
        0 && 1
    \end{pmatrix}
    \mathbf{V}_{\text{\footnotesize LO}}^{-1}(N)\,.
\end{align}
The operators $\widetilde{\mathbf{R}}_k$ encode the iterative procedure:
\begin{align}
    \label{eq:Rtilde_operators}
    \widetilde{\mathbf{R}}_k = \mathbf{R}_k + \sum_{i=1}^{k-1} \mathbf{R}_{k-i} \mathbf{U}_i
    \,.
\end{align}
The framework above can be implemented in at least three different ways, corresponding to the three available evolution modes implemented in \verb|PEGASUS|~\cite{Vogt:2004ns}. However, none of them provides a fully exponentiated solution and only one leads to a closed expression. In the following, we briefly discuss the realization of these three modes at NLO.

\bigskip

A first, natural approach consists in neglecting all the terms associated to $\mathbf{P}_{k\geq 2}$ and $b_{k\geq 2}$ in Eq.~\eqref{eq:singlet_EOevo_nMloR}. This choice reproduces exactly the NLO equation of Eq.~\eqref{eq:singlet_EOevo_nloR} and it has been widely used in parton density analyses~\cite{Lai:1996mg,Martin:1995ws}.
Moreover, it simplifies the expressions for the $\mathbf{R}$-operators to\cite{Blumlein:1997em}:
\begin{align}
\label{eq:imodev_1}
    &\mathbf{R}_k^{\text{\footnotesize NLO}} = (-b_1)^{k-1}\mathbf{R}_1,
\end{align}
which, in turn, reflects on the operators $\widetilde{\mathbf{R}}_k$ and the matrices $\mathbf{U}_k$, both depending on the sole $\mathbf{R}_1$. In the end, the solution is obtained iteratively. It can be regarded as the iterative counterpart of the result obtained in the previous Section in Eq.~\eqref{eq:singlet_EO_solNLO}. This fact is highlighted by the implicit presence of Gauss Hypergeometric functions in the definition of the operator $\mathbf{U}$, which reflects the functional form found in Eqs.~\eqref{eq:ingredients_NLO_funs}. For example, consider the projection $\mathbf{e}_- \mathbf{U} \mathbf{e}_+$ and then apply the definition of $\widetilde{\mathbf{R}}_k$ together with Eq.~\eqref{eq:imodev_1}. We obtain:
\begin{align}
    \label{eq:proj_example}
    &\mathbf{e}_- \mathbf{U}\left(N;a_S\right) \mathbf{e}_+
    =
    \mathbf{e}_- 
    \mathbf{R}_1
    \Bigg[
    -\frac{\log{(1+b_1\,a_S)}}{b_1}
    -\frac{a_S}{1+\Delta_0}
    {}_2 F_1\left(1,1+\Delta_0;2+\Delta_0;-b_1\,a_S\right)
    -
    \notag \\
    &\quad-
    \sum_{i=1}^\infty 
    a_S^{1+i}
    \left(
    \frac{{}_2 F_1\left(1,1+i;2+i;-b_1\,a_S\right)}{1+i}
    +
    \frac{{}_2 F_1\left(1,1+i+\Delta_0;2+i+\Delta_0;-b_1\,a_S\right)}{1+i+\Delta_0}
    \right)
    \mathbf{U}_i
    \Bigg]
    \mathbf{e}_+,
\end{align}
and similarly for the other projections. Remarkably, the differences between the iterative result devised above and the closed exponentiated solution of Eq.~\eqref{eq:singlet_EO_solNLO} do not originate from the different choice of the initial ansatz, but rather on the methodology adopted to solve the evolution of the operator $\mathbf{U}$. In fact, at NLO it solves:
\begin{align}
    \label{eq:Umatrices_diffeq}
    \frac{\partial \mathbf{U}\left(N; a_S\right)}{\partial a_S}
    =
    -\frac{1}{a_S}\comm{\mathbf{R}_0(N)}{\mathbf{U}\left(N; a_S\right)}
    -\frac{1}{1+b_1 a_S}
    \mathbf{R}_1(N) \mathbf{U}\left(N; a_S\right),
\end{align}
which is a Heisenberg-like evolution equation. 
Setting $\mathbf{U} = e^{\mathbf{S}_0} \widetilde{\mathbf{U}} e^{-\mathbf{S}_0}$ and then solving for $\widetilde{\mathbf{U}}$ would lead to the \emph{same} result obtained in Eq.~\eqref{eq:singlet_EO_solNLO}. This explicitly shows that $\mathbf{U}$ is a function of both $a_S$ and $a_0$, somehow in contradiction with the expansion of Eq.\eqref{eq:U_exp}.

\bigskip

Different solutions can be obtained by applying the NLO approximation already at the level of the “Hamiltonian". The evolution operator does not obey anymore Eq.\eqref{eq:singlet_EOevo_nloR} as in the previous cases, but rather:
\begin{align}
    \label{eq:singlet_EOevo_nloR_appr}
    \frac{\partial}{\partial a_S} \mathbf{E}\left(N;a_S\right)
    =
    \left(
    -\frac{1}{a_S} \mathbf{R}_0(N)
    -\mathbf{R}_1(N)
    \right)
    \mathbf{E}\left(N;a_S\right).
\end{align}
In such cases the solution is necessarily accompanied by larger theoretical uncertainties, as the approximations made at the level of the evolution equation add to the theoretical errors associated to the matrix nature of the singlet sector. Nevertheless, iterative algorithms can minimize this effect. 
In this regard, the analogue of Eq.\eqref{eq:imodev_1} is:
\begin{align}
    \label{eq:imodev_2}
    &\mathbf{R}^{\text{\footnotesize NLO}}_{k\geq 2} = 0
\end{align}
which implies $\widetilde{\mathbf{R}}_k = \mathbf{R}_1 \mathbf{U}_{k-1}$ for $k\geq 2$. This choice leads to a second iterative solution, in which the projections of the operator $\mathbf{U}$ simplify to:
\begin{align}
    \label{eq:proj_example_2}
    &\mathbf{e}_- \mathbf{U}\left(N;a_S\right) \mathbf{e}_+
    =
    \mathbf{e}_-
    \mathbf{R}_1
    \left[
    -a_S \left(1 + \frac{1}{\Delta_0 + 1}\right)
    -
    \sum_{k\geq 2} a_S^k \left(\frac{1}{k} + \frac{1}{\Delta_0+k}\right)
    \mathbf{U}_{k-1}
    \right]
    \mathbf{e}_+,
\end{align}
and analogously for the other projections. 

\bigskip

Finally, the expansion in the strong coupling can be further exploited and applied to the final solution. This leads to the closed result:
\begin{align}
    \label{eq:singlet_truncatedSol_NLO}
    &\mathbf{E}_{\text{\footnotesize tr.}}^{\text{\footnotesize NLO}}\left(N;a_0,a_S\right)
    =
    e^{h_1(a_0,a_S) \mathbf{R}_0(N)}
    +
    a_S \mathbf{U}_1(N)
    \,
    e^{h_1(a_0,a_S) \mathbf{R}_0(N)}
    -
    a_0 \,
    e^{h_1(a_0,a_S) \mathbf{R}_0(N)}
    \mathbf{U}_1(N)
\end{align}
which is sometimes referred to as the “truncated" solution at NLO and, in the following, we will adopt this nomenclature.
Such strategy is extremely popular and it is the foundation of most code implementations based on analytic solutions of DGLAP equations in Mellin space. 
Due to its simpler structure, the approximation in Eq.~\eqref{eq:singlet_truncatedSol_NLO} is particularly suitable for code implementation~\cite{Vogt:2004ns,Moffat:2021dji,NNPDF:2024djq}.
However, the violation to the original differential equation introduced by Eq.~\eqref{eq:singlet_truncatedSol_NLO} is unfortunately non zero at the input scale, as we will show in Section~\ref{sec:viol}.

\begin{figure}[t]
\includegraphics[width=.9\textwidth]{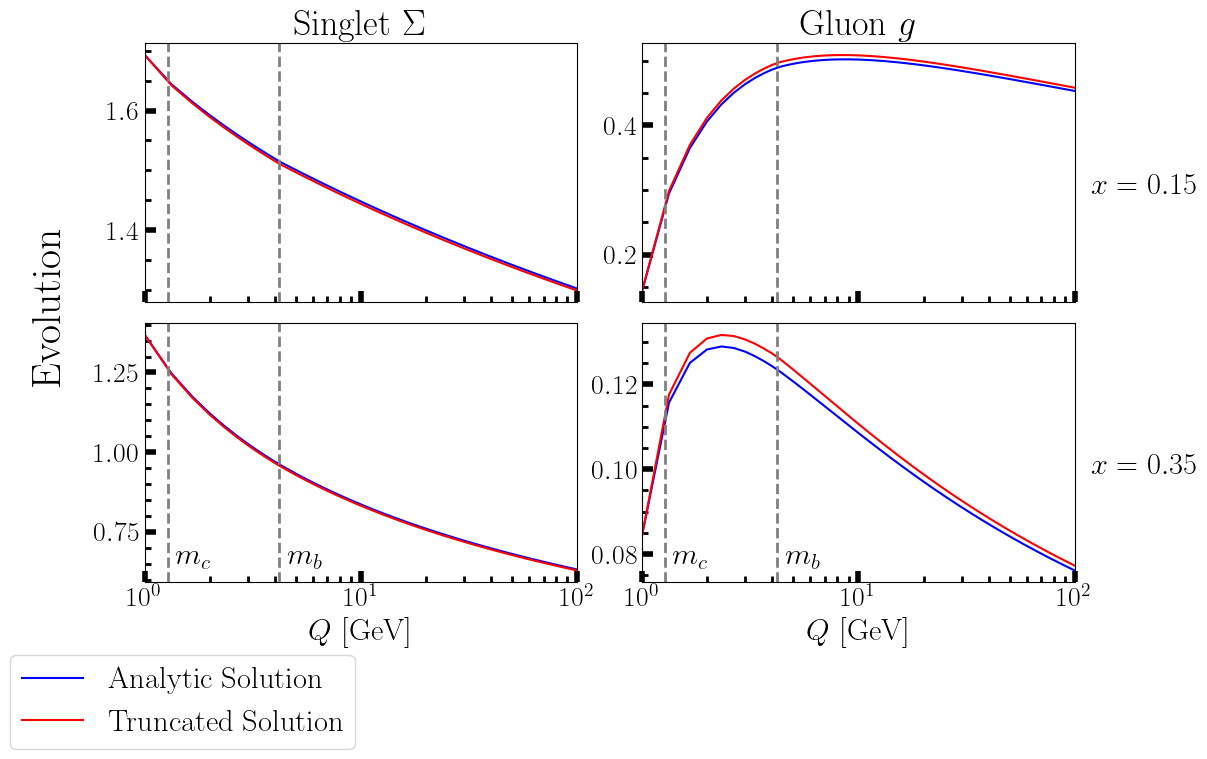}
\caption{Comparison of the action of the Evolution Operators $\mathbf{E}^{\text{\footnotesize NLO}}$ of Eq.~\eqref{eq:singlet_EO_solNLO} (Analytic Solution) and $\mathbf{E}_{\text{\footnotesize tr.}}^{\text{\footnotesize NLO}}$ of Eq.~\eqref{eq:singlet_truncatedSol_NLO} (Truncated Solution) on the simple proton model described in Eqs.~\eqref{eq:input_param}. 
The distributions in $x$-space are defined in Eq.~\eqref{eq:ratio}.
The left panels refer to the flavor-singlet quark distributions $\Sigma = u_V + \,d_V + 6\,q_{\text{\footnotesize sea}}$, while the right panels are dedicated to the gluon distributions $g$. 
The thresholds at charm and bottom quarks are shown.
\label{fig:evo}}
\end{figure}

The differences between the two closed solutions $\mathbf{E}^{\text{\footnotesize NLO}}$ of Eq.~\eqref{eq:singlet_EO_solNLO} and $\mathbf{E}_{\text{\footnotesize tr.}}^{\text{\footnotesize NLO}}$ of Eq.~\eqref{eq:singlet_truncatedSol_NLO} can be addressed by acting with these two Evolution Operators on the same input functions and comparing the results. We adopt a simple model for proton PDFs, where the only requirement is that the sum rules associated to momentum conservation and valence quark numbers are satisfied. This simple model is:
\begin{subequations}
\label{eq:input_param}
    \begin{align}
    &u_V(N;Q_0) = 2 \frac{B(\alpha_u+N,\beta_u+1)}{B(\alpha_u+1,\beta_u+1)}
    \\
    &d_V(N;Q_0) = \frac{B(\alpha_d+N,\beta_d+1)}{B(\alpha_d+1,\beta_d+1)}
    \\
    &g(N;Q_0) = \gamma_g \,  B(\alpha_g+N,\beta_g+1)
    \\
    &q_{\text{\footnotesize sea}}(N;Q_0) = \gamma_{\text{\footnotesize sea}}\,
    B(\alpha_{\text{\footnotesize sea}}+N,\beta_{\text{\footnotesize sea}}+1)
    \quad\text{with: }
    \gamma_{\text{\footnotesize sea}} = \frac{1-u_V(2;Q_0)-d_V(2;Q_0)-g(2,Q_0)}{ 6\,B(\alpha_{\text{\footnotesize sea}}+2,\beta_{\text{\footnotesize sea}}+1)}
    \end{align}
\end{subequations}
where $B$ is the Euler Beta function. In total, there are 9 parameters which should be fixed with a fit at the input scale $Q_0$, that we assume to be $1$ GeV. Although phenomenological applications are the ultimate goal, in this work we are primarily interested in comparing different treatments of the theoretical uncertainties in the solution of the DGLAP equations.
Thus, in the following the parameters of the simple model of Eq.~\eqref{eq:input_param} will be fixed to some sensible choice and  and we defer to future studies the actual phenomenological analysis on experimental data. 
The comparison is presented in Fig.~\ref{fig:evo} for the singlet sector at two different values of $x$, where we consider the distributions in $x$-space obtained as:
\begin{align}
    \label{eq:ratio}
    &\mathbf{q}^{\text{\footnotesize sol}}_S(x,Q) =
    \int \frac{d N}{2\pi i} x^{-N}
    \,
    \mathbf{E}^{\text{\footnotesize sol}}\left(N;a_0,a_S(Q)\right) \, 
    \mathbf{q}_S(x,Q_0)
    \quad\text{with: }
    \mathbf{q}_S = \begin{pmatrix}
        \Sigma \\
        g
    \end{pmatrix}.
\end{align}
The label “sol" refers to either one of the solutions of Eqs.~\eqref{eq:singlet_EO_solNLO} and Eq.~\eqref{eq:singlet_truncatedSol_NLO}. 
The inverse Mellin transform is performed numerically, by using the same routine adopted in other codes~\cite{Vogt:2004ns} based on analytic approaches to DGLAP evolution.
Interestingly, from Fig.~\ref{fig:evo} we observe that the different evolution affects more significantly the gluon distribution $g$, compared to the flavor-singlet distribution $\Sigma$.
Such differences are nevertheless higher order and the curves look very similar to each other. 
In fact, the size of the discrepancies is directly related to the different size of the theoretical errors and a more refined test is thus necessary to properly address them. This will be the subject of the next Section.

\section{Theoretical uncertainties and Violation of the Evolution Equations\label{sec:viol}}

In order to study the violation induced by the approximated solution $\mathbf{E}^{\text{\footnotesize sol}}$, we introduce the Violation Operator:
\begin{align}
    \label{eq:violation_operator}
    &\mathbf{V}^{\text{\footnotesize sol}}\left(N;a_0,a_S(Q)\right)=
    \frac{\partial \mathbf{E}^{\text{\footnotesize sol}}\left(N;a_0,a_S(Q)\right)}{\partial \log{Q^2}}
    - \left(a_S(Q) \mathbf{P}_0(N) + 
    a_S^2(Q) \mathbf{P}_1(N)
    \right)
    \mathbf{E}^{\text{\footnotesize sol}}\left(N;a_0,a_S(Q)\right).
\end{align}
Similarly to Eq.~\eqref{eq:ratio}, we define the distributions associated to the violation as:
\begin{align}
    \label{eq:ratio_viol}
    &\Delta\mathbf{q}^{\text{\footnotesize sol}}_S(x,Q) =
    \int \frac{d N}{2\pi i} x^{-N}
    \,
    \mathbf{V}^{\text{\footnotesize sol}}\left(N;a_0,a_S(Q)\right) \, 
    \mathbf{q}_S(x,Q_0)
    \quad\text{with: }
    \Delta\mathbf{q}_S = \begin{pmatrix}
        \Delta\Sigma \\
        \Delta g
    \end{pmatrix}
\end{align}
The more the solution is accurate, the more the violation above is smaller. 
We can explicitly compute the violations induced by the Analytic Solution of Eq.~\eqref{eq:singlet_EO_solNLO} and the Truncated Solution of Eq.~\eqref{eq:singlet_truncatedSol_NLO}. They are, respectively:
\begin{subequations}
\begin{align}
    \label{eq:analyticSOL_violation}
    &\mathbf{V}^{\text{\footnotesize NLO}}\left(N;a_0,a_S\right)=
    a_S^2 \beta_0 e^{h_1(a_0,a_S)\mathbf{R}_0(N)}
    \Bigg(
    \comm{\mathbf{E}_{\text{\footnotesize int}}\left(N;a_0,a_S\right)}{ e^{-h_1(a_0,a_S)\mathbf{R}_0(N)}\comm{\mathbf{R}_1(N)}{e^{h_1(a_0,a_S)\mathbf{R}_0(N)}}}
    \notag \\
    &\quad-
    \frac{\sinh{\left(
    h_1(a_0,a_S)\Delta_0(N)
    \right)}}{\Delta_0(N)}
    \mathbf{E}_{\text{\footnotesize int}}\left(N;a_0,a_S\right)
    e^{-h_4(\Delta_0(N),a_0,a_S)} 
    \comm{\comm{\mathbf{R}_0(N)}{\mathbf{R}_1(N)}}{
    e^{h_4(\Delta_0(N),a_0,a_S)}}
    \Bigg)
    \\
    \label{eq:truncatedSOL_violation}
    &\mathbf{V}_{\text{\footnotesize tr.}}^{\text{\footnotesize NLO}}\left(N;a_0,a_S\right)=
    a_S^3 \beta_1 \mathbf{R}_1(N) e^{h_1(a_0,a_S) \mathbf{R}_0(N)}
    -a_S^2 \beta_0 \left(
    a_S \mathbf{U}_1(N)
    \,
    e^{h_1(a_0,a_S) \mathbf{R}_0(N)}
    -
    a_0 \,
    e^{h_1(a_0,a_S) \mathbf{R}_0(N)}
    \mathbf{U}_1(N)
    \right)
\end{align}
\end{subequations}
where the operator $\mathbf{E}_{\text{\footnotesize int}} = e^{-h_1 \, \mathbf{R}_0} \mathbf{E}^{\text{\footnotesize NLO}}$ is the product of the last three factors in Eq.~\eqref{eq:singlet_EO_solNLO}. From these expressions and from the fact that all the functions $h_i$ vanishes at the input scale,
it follows that $\mathbf{V}^{\text{\footnotesize NLO}}$ is correctly zero in the limit $a_S \to a_0$, while $\mathbf{V}_{\text{\footnotesize tr.}}^{\text{\footnotesize NLO}}$ tends to $a_0^3 \beta_1 \mathbf{R}_1$. This is clearly a N$^3$LO violation, hence higher order with respect to the accuracy of the solution. However, it still is a matter of concern, as it spoils the assumption that the Evolution Operator is unity at the input scale.
\begin{figure}[t]
\includegraphics[width=1\textwidth]{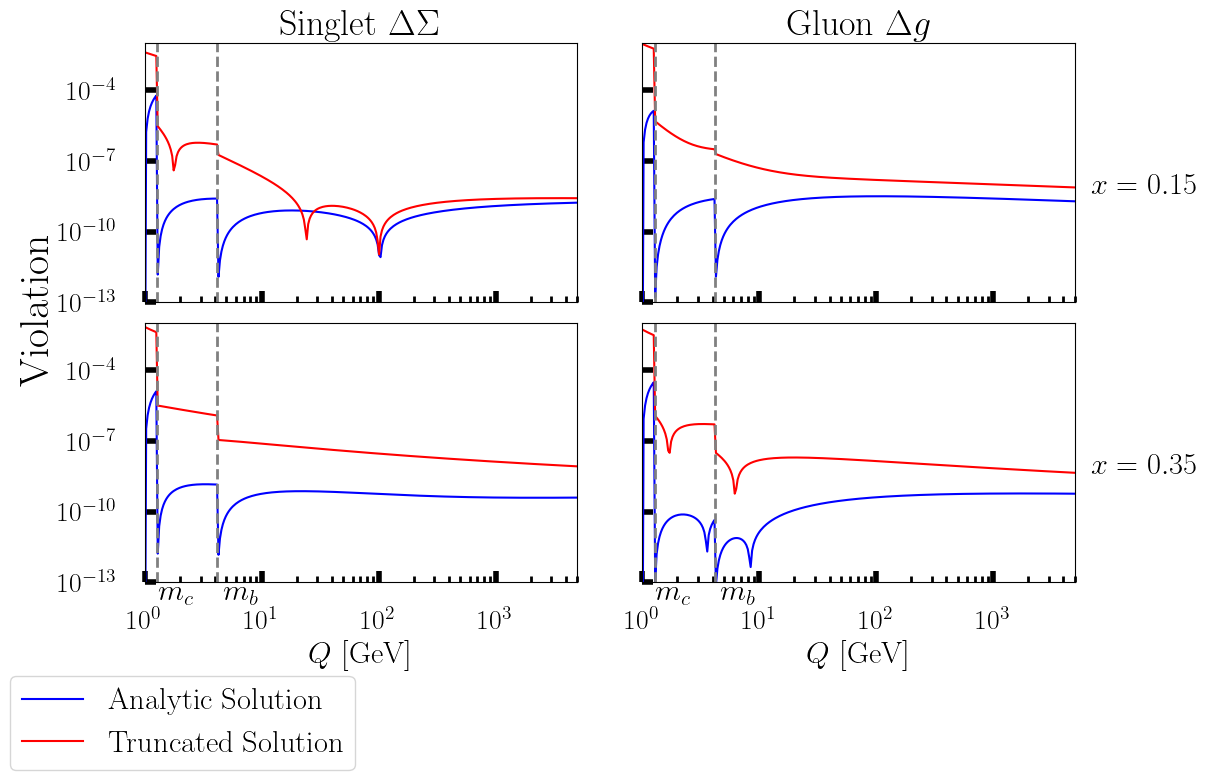}
\caption{Comparison of the action of the Violation Operators $\mathbf{V}^{\text{\footnotesize NLO}}$ of Eq.~\eqref{eq:analyticSOL_violation} (Analytic Solution) and $\mathbf{V}_{\text{\footnotesize tr.}}^{\text{\footnotesize NLO}}$ of Eq.~\eqref{eq:violation_operator} (Truncated Solution) on the simple proton model described in Eqs.~\eqref{eq:input_param} for $1$ Gev $\geq Q \geq 5000$ GeV.
Quark thresholds are indicated by the vertical dashed lines in the upper panel. 
The violation induced by the Analytic Solution of is systematically smaller than the violation introduced by the truncated solution, by several orders of magnitude at $Q \approx Q_0$. The contribution of the top quark is not considered.}
\label{fig:viol}
\end{figure}

\section{Logarithmic Accuracy}

The energy dependence of the analytic solutions obtained so far has been parametrized in terms of the strong coupling $a_S = \alpha_S/(4\pi)$, evaluated at the energy scale and at scale $Q$, where clearly, $a_S(Q) \leq a_S(Q_0)$ on the support of Eq.~\eqref{eq:DGLAP_eq}. These two values are commonly obtained by integrating numerically the Renormalization Group (RG) evolution for the strong coupling:
\begin{align}
    \label{eq:aS_evolution}
    \frac{d}{d \log{Q^2}} a_S(Q) = \beta\left(a_S(Q) \right),
\end{align}
where the beta function is truncated at the desired order in $a_S$. The coefficients of the expansions of $\beta(a_S)$ are currently known up to $5$ loops~\cite{Baikov:2016tgj} but usually only the orders required for the specific analysis are kept. For instance, the $h_i$ functions calculated in Eqs.~\eqref{eq:ingredients_NLO_funs} have been obtained truncating the expansion for the $\beta$ function at NLO and hence the solution of Eq.~\eqref{eq:aS_evolution} has to be consistently considered at the same perturbative order. 

Depending on the purpose of the analysis, it might be worth to solve analytically also the RG evolution for $a_S$. 
The boundary condition is usually chosen to be the most precise experimental measure available~\cite{ATLAS:2023lhg} for $\alpha_S$, taken at the mass $m_Z$ of the $Z$ boson. 
Couplings at different scales are then expressed in terms of $a_Z = a_S(m_Z)$.
Although an exact solution can only be obtained at LO, it is possible to obtain good approximations of the numeric integrations by using the following expansion:
\begin{align}
    \label{eq:aS_analytic}
    a_S(Q) = \frac{a_Z}{1-\lambda} 
    \left(
    1 - a_Z \frac{\beta_1}{\beta_0} \frac{\log{(1-\lambda)}}{1-\lambda}
    - a_Z^2 \frac{\beta_1^2}{\beta_0^2}\frac{1}{(1-\lambda)^2}
    \left(
    \log{(1-\lambda)} - \log^2{(1-\lambda) + \left(1 - \frac{\beta_0 \beta_2}{\beta_1}\right)\lambda}
    \right)
    + \dots
    \right)
\end{align}
where $\lambda = 2 a_Z \beta_0 \log{(m_Z/Q)}$.
At the input scale, where the logarithms size is the largest possible and the errors are maximized, the first two terms of this expansion produce a discrepancy about $2\%$ with the numerical solution at NLO, decreasing to about $1\%$ if the first three terms are compared with the NNLO numeric integration. 

\bigskip

The approximations introduced to get to Eq.~\eqref{eq:aS_analytic} can be iterated to express the coupling at the input scale $a_0$ in terms of the coupling $a_S(Q)$. This approach would effectively result in an expansion in (inverse) powers of the logarithms $L = \log{(Q/Q_0)}$ and its accuracy would degrade as the energy $Q$ increases. Substituting it in Eqs.~\eqref{eq:ingredients_NLO_funs}, we obtain the following expansions for the functions $h_i$:
\begin{subequations}
\label{eq:ingredients_NLO_funs_logexp}
    \begin{align}
    &h_1(a_0,a_S) = f_1(\lambda) + \frac{1}{L} f_2^{(1)}(\lambda) + \mathcal{O}\left(\frac{1}{L^2}\right);
    \\
    &h_2(a_0,a_S) = \frac{1}{L} f_2^{(2)}(\lambda) + \mathcal{O}\left(\frac{1}{L^2}\right);
    \\
    &h_3(\Delta,a_0,a_S) = \frac{1}{L} f_3(\Delta,\lambda) + \mathcal{O}\left(\frac{1}{L^2}\right);
    \\
    &h_4(\Delta,a_0,a_S) = \frac{1}{L} f_4(\Delta,\lambda) + \mathcal{O}\left(\frac{1}{L^2}\right),
    \end{align}
\end{subequations}
where this time $\lambda = 2 a_S(Q) \beta_0 L$ and the functions $f_i$ are:
\begin{subequations}
\label{eq:fi_funs}
    \begin{align}
    &f_1(\lambda) = -\log{(1-\lambda)};
    \\
    &f_2^{(1)}(\lambda) = -\frac{1}{2 \beta_0} \frac{\beta_1}{\beta_0} \frac{\lambda}{1-\lambda}\log{(1-\lambda)};
    \\
    &f_2^{(2)}(\lambda) =
    \frac{1}{2 \beta_0} \frac{\lambda^2}{1-\lambda};
    \\
    &f_3(\Delta,\lambda) = 
    -\frac{1}{4 \beta_0}\lambda
    \left(
    \frac{2}{(1-\Delta^2)}\frac{1}{1-\lambda}
    +\frac{1}{\Delta}\left(
    \frac{(1-\lambda)^\Delta}{1+\Delta}
    -
     \frac{(1-\lambda)^{-\Delta}}{1-\Delta}
    \right)
    \right);
    \\
    &f_4(\Delta,\lambda) = 
    -\frac{1}{4 \beta_0}\frac{\lambda}{1-\lambda}\frac{1}{\Delta^2}
    \left(-\frac{2 \left(1 - (1-\Delta^2)\lambda\right)}{1-\Delta^2}
    +
    \frac{(1-\lambda)^\Delta}{1+\Delta}
    +
     \frac{(1-\lambda)^{-\Delta}}{1-\Delta}
     \right).
    \end{align}
\end{subequations}
The expressions in Eq.~\eqref{eq:ingredients_NLO_funs_logexp} can be legitimately regarded as Next-Leading-Log (NLL) accurate. The determination of the next log-order, proportional to $L^{-2}$, would systematically require the use of the QCD beta function at NNLO. 
The functions $f_i$ are singular in $\lambda=1$, which sets the limit where the expansion in powers of logarithms stops to converge. 

The application of this strategy to the solution of the DGLAP evolution leads to the NLL expression for the Evolution Operator determined in Eq.~\eqref{eq:singlet_EOevo_nlo}. In this regard, we can now express it in terms of the original splitting kernels $\mathbf{P}$ instead of the $\mathbf{R}$ operators introduced in Eq.~\eqref{eq:R_operators}. The only contribution that has to be re-arranged is the second matrix exponential in Eq.~\eqref{eq:singlet_EOevo_nlo}, as its exponent is related to the sum of $\mathbf{P}_0$ and $\mathbf{P}_1$. However, at NLL it can be equivalently written as:
\begin{align}
    \label{eq:recaste_E2}
    &\text{exp}\left(
    h_2(a_0,a_S) \mathbf{R}_1
    \right)
    =
    \text{exp}\left(
    \frac{1}{L}f_2^{(2)}(\lambda) \left(\frac{1}{\beta_0} \mathbf{P}_1- \frac{\beta_1}{\beta_0^2}\mathbf{P}_0\right)
    +\mathcal{O}\left(\frac{1}{L^2}\right)
    \right)
    =
    \notag \\
    &\quad
    \text{exp}\left(
    -\frac{1}{L}\frac{\beta_1}{\beta_0^2}f_2^{(2)}(\lambda)\mathbf{P}_0
    \right)
    \text{exp}\left(
    \frac{1}{L}f_2^{(2)}(\lambda) \frac{1}{\beta_0} \mathbf{P}_1
    \right)
    \, \left(
    1 + \mathcal{O}\left(\frac{1}{L^2}
    \right)\right)
\end{align}
where in the last step we have used the Zassenhaus formula of Eq.~\eqref{eq:Zassenhaus}. The example above shows how easy is to compute the leading contributions at a given log-order, as the size of every term is uniquely and completely fixed.
In conclusion, the NLL Evolution Operator reads as:
\begin{align}
    \label{eq:singlet_EO_solNLL}
    &\mathbf{E}^{\text{\footnotesize NLL}}\left(N; Q_0,Q\right)
    =
     \text{exp}\left(
     \left(
     \widetilde{f}_1(\lambda) + \frac{1}{L} \widetilde{f}_2^{\;(1)}(\lambda)
     \right) 
     \mathbf{P}_0(N)
     \right)
     \text{exp}\left(\frac{1}{L}\widetilde{f}_2^{\;(2)}(\lambda) \mathbf{P}_1(N)\right)
    \notag \\
     &\quad\times
     \text{exp}\left(\frac{1}{L}\widetilde{f}_3\left(\frac{\widetilde{\Delta}_0(N)}{\beta_0},\lambda\right)  \comm{\mathbf{P}_0(N)}{\mathbf{P}_1(N)}\right)
     \text{exp}\left(\frac{1}{L}\widetilde{f}_4\left(\frac{\widetilde{\Delta}_0(N)}{\beta_0},\lambda\right) \comm{\mathbf{P}_0(N)}{ \comm{\mathbf{P}_0(N)}{\mathbf{P}_1(N)}}\right).
\end{align}
where $\widetilde{\Delta}_0$ is the difference of the eigenvalues of $\mathbf{P}_0$ and the $\widetilde{f}_i$ functions are:
\begin{subequations}
\label{eq:fitilde_funs}
    \begin{align}
    &\widetilde{f}_1(\lambda) = \frac{1}{\beta_0}f_1(\lambda);
    \\
    &\widetilde{f}_2^{\;(1)}(\lambda) = \frac{1}{\beta_0} \left(
    f_2^{\;(1)}(\lambda) - b_1 f_2^{\;(2)}(\lambda)
    \right);
    \\
    &\widetilde{f}_2^{\;(2)}(\lambda) =
     \frac{1}{\beta_0}f_2^{\;(2)}(\lambda);
    \\
    &\widetilde{f}_3(\Delta,\lambda) = 
    \frac{1}{\beta_0^2}f_3(\Delta,\lambda);
    \\
    &\widetilde{f}_4(\Delta,\lambda) = 
    \frac{1}{\beta_0^3}f_4(\Delta,\lambda).
    \end{align}
\end{subequations}
Notice that the functions $\widetilde{f}_1$ and $\widetilde{f}_2$ precisely correspond to the NLL approximation of the integrals $\int_{Q_0}^Q d (\log{\mu^2}) a_S(\mu)$ and $\int_{Q_0}^Q d (\log{\mu^2}) a^2_S(\mu)$, respectively.

\bigskip

The solution presented in Eq.~\eqref{eq:singlet_EO_solNLL} is remarkable for many aspects. 
First of all, while it is possible to give different interpretations of the label “NLO" and produce different analytic expressions as testified by Eqs.~\eqref{eq:singlet_EO_solNLO} and~\eqref{eq:singlet_truncatedSol_NLO}, the meaning of NLL is unambiguous and it specifically refers to a result in which all the relevant logarithms are exponentiated and organized in descending powers. Such expressions are often at the core of many observables and suited for a wide range of applications, from event-shape observables to transverse momentum dependent cross sections~\cite{Catani1989,Catani:1991kz,Catani:1992ua,Bozzi:2010xn,Gross:2022hyw}, where the relevant log-orders are usually obtained within the framework of resummation.
In particular, parton distributions are used as inputs in Transverse Momentum Dependent (TMD) densities, where it is common practice to take into account the log-ordering in terms of logarithms of transverse distance~\cite{Bacchetta:2022awv,Boglione:2023duo,Moos:2023yfa}.
Although it is beyond the purpose of this paper to discuss about the correct implementation of the log counting within the TMD factorization
and there is a dedicated recent literature on this subject~\cite{Gonzalez-Hernandez:2022ifv,Gonzalez-Hernandez:2023iso,Aslan:2024nqg}, the solution obtained in Eq.~\eqref{eq:singlet_EO_solNLL} would be particularly suited for applications in the TMD case and for simultaneous extractions of collinear parton densities together with their TMD counterparts. 
More generally, studying the effects of the collinear functions in TMD observables is triggering the interest of the TMD community~\cite{Gonzalez-Hernandez:2022ifv,Gonzalez-Hernandez:2023iso,Boglione:2022nzq,Barry:2023qqh} and most modern analyses are also devoted to this topic.

Notice that also the truncated solution of Eq.~\eqref{eq:singlet_truncatedSol_NLO} can be expanded in powers of $L$, but such logarithms would not be exponentiated and the resulting approximation would not properly resum all the NLL contributions to the parton distributions. Therefore, Eq.~\eqref{eq:singlet_EO_solNLL} is the only NLL accurate result (in the sense specified above) for the singlet sector of the DGLAP evolution.
Moreover, it is completely based on a analytic approach: every ingredient in $\mathbf{E}^{\text{\footnotesize NLL}}$ has been explicitly computed analytically. 
Finally, we stress once again that the expansion in powers of $1/L$ is extremely transparent and all the neglected terms are assigned with a well defined scaling. 
As a consequence, the Violation Operator of Eq.~\eqref{eq:violation_operator} associated to the NLL solution is certainly of order $1/L^2$, as well as the effect of the perturbative hysteresis~\cite{Bertone:2022sso}. 

\begin{figure}[t]
\includegraphics[width=.9\textwidth]{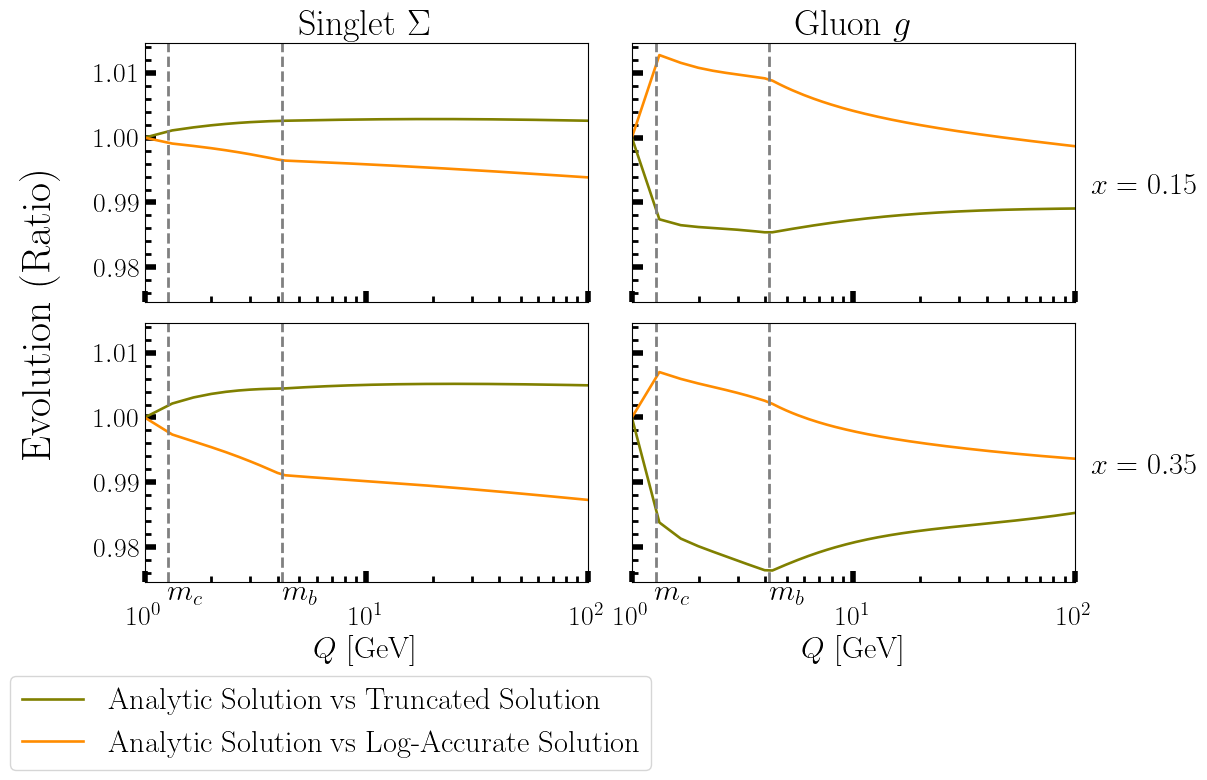}
\caption{Comparison of the action of the Evolution Operators $\mathbf{E}^{\text{\footnotesize NLO}}$ of Eq.~\eqref{eq:singlet_EO_solNLO} (Analytic Solution), $\mathbf{E}_{\text{\footnotesize tr.}}^{\text{\footnotesize NLO}}$ of Eq.~\eqref{eq:singlet_truncatedSol_NLO} (Truncated Solution) and $\mathbf{E}^{\text{\footnotesize NLL}}$ (Log-Accurate Solution) on the simple proton model described in Eqs.~\eqref{eq:input_param}. 
The lines refer to ratios of distributions in $x$-space defined in Eq.~\eqref{eq:ratio}. 
Green lines compare the Analytic Solution and the Truncated Solution, while the orange lines compare the Analytic Solution and the Log-Accurate Solution.
The left panels refer to the flavor-singlet quark distributions $\Sigma = u_V + \,d_V + 6\,q_{\text{\footnotesize sea}}$, while the right panels are dedicated to the gluon distributions $g$. 
The thresholds at charm and bottom quarks are shown.
\label{fig:evoFULL}}
\end{figure}

Despite its appealing features, the solution in Eq.~\eqref{eq:singlet_EO_solNLL} also has disadvantages. 
In particular, being an approximation of the (already approximated) NLO result of Eq.~\eqref{eq:singlet_EOevo_nlo}, it is less precise when compared with the exact numerical solution of the DGLAP evolution. 
Therefore, given the differential equation of Eq.~\eqref{eq:singlet_DGLAP}, the choice of which solution to adopt among the three in Eqs.~\eqref{eq:singlet_EO_solNLO},~\eqref{eq:singlet_truncatedSol_NLO} and ~\eqref{eq:singlet_EO_solNLL} ultimately depends on the problem at hand. 
A comprehensive comparison for all the analytic solutions discussed in this work is shown in Fig.~\ref{fig:evoFULL}.

\section{Conclusions}

The DGLAP evolution equations are at the core of the research on strong interactions  and in the analysis of hadronic processes, as they allow to predict the energy dependence of parton PDFs. 
As modern experiments demand ever increasing precision, the quest for accurate solutions becomes imperative. While numerical integration of Eqs.~\eqref{eq:DGLAP_eq} offers one avenue, exploring analytic approaches proves at least equally fruitful. Explicit analytic expressions not only afford a clearer understanding of the evolution's effects but also yield a more refined assessment of theoretical uncertainties. Such considerations are particularly pertinent in the singlet sector, where obtaining a closed analytic form is elusive, necessitating reliance on approximations.

In this study, we investigate three distinct analytical closed solutions at NLO, comparing and contrasting their merits. We introduce a mathematical formalism for handling evolution in two dimensions and leverage it to derive a novel solution aimed at closely approximating the exact numerical solution while preserving the expected mathematical properties, such as exponentiation (as seen in Eq.~\eqref{eq:singlet_EO_solNLO}). This Analytical Solution exhibits systematic precision superior to the commonly employed Truncated Solution (Eq.~\eqref{eq:singlet_truncatedSol_NLO}), widely utilized in phenomenological analyses rooted in analytical methodologies. However, it entails a more intricate structure involving four matrix exponentials and non-trivial functions. Notably, despite its complexity, it requires only the diagonalization of three matrices associated with the splitting kernels up to NLO and their commutator. Additionally, we present the Log-Accurate Solution at NLL (Eq. ~\eqref{eq:singlet_EO_solNLL}), which, while less accurate than the NLO solutions, provides a cleaner estimation of theoretical uncertainties. In fact, all the neglected terms, as well as the size of the violation of the original differential equation, scale as $1/L^2$.
Ultimately, the choice among these solutions hinges on the specific aims of the analysis. For precision and mathematical elegance, the Analytic Solution (Eq.~\eqref{eq:singlet_EO_solNLO}) may be preferred, while the Log-Accurate Solution (Eq.~\eqref{eq:singlet_EO_solNLL}) might find its purpose in scenarios prioritizing error control or integration into the resummation formalisms, including the TMD case. Conversely, the Truncated Solution (Eq.~\eqref{eq:singlet_EO_solNLL}) may be favored for its simplicity, facilitating rapid code implementations.

In summary, I provide a comprehensive study of various analytical approaches to the singlet DGLAP evolution, presenting explicit computations of novel analytical results alongside an assessment of their advantages and drawbacks. This work lays the ground for enhancing existing numerical codes based on analytical DGLAP solutions and for the development of refined tools, promising advancements in the realm of Parton Distributions and the understanding of hadronic processes.

 \begin{acknowledgments}
     I would like to thank Nobuo Sato for useful discussions and for providing the code routine for the inverse Mellin transform. I am also grateful to Alberto Accardi, Patrick Barry,  Marielena Boglione and Matteo Cerutti for their valuable suggestions.
     This work is supported in part by the US Department of Energy (DOE) Contract No.~DE-AC05-06OR23177, under which Jefferson Science Associates, LLC operates Jefferson Lab.
 \end{acknowledgments}

\appendix 

\section{Results for matrix representations in two dimensions\label{app:mat2x2}}

In this Section, we collect some useful results holding in $2$-dimensions. 
We introduce the following notation for the nested $n$-commutator of the matrix $\mathbf{A}$ with the matrix $\mathbf{B}$:
\begin{align}
    \label{eq:n-comm}
    &\comm{\left(\mathbf{A}\right)^n}{\mathbf{B}}
    =
    \comm{\mathbf{A}}{\comm{\mathbf{A \big[\dots 
    }}{\mathbf{B}},\dots\big] },
\end{align}
where the matrix $\mathbf{A}$ appears $n$ times. 
The main simplifications occurring in $2$ dimensions are consequences of the following Lemma:
\begin{lemma}
\label{lemma:principal}
Let $\mathbf{A}$ and $\mathbf{B}$ be $2$ dimensional diagonalizable matrices and let $K(\mathbf{A}) = \text{Tr}(\mathbf{A})^2 - 4 \text{Det}(\mathbf{A})$.  
If the commutator $\comm{\mathbf{A}}{\mathbf{B}}$ is diagonalizable, then every $n$-commutator of $\mathbf{A}$ with $\mathbf{B}$ is proportional either to $\comm{\mathbf{A}}{\mathbf{B}}$ or to $\comm{\mathbf{A}}{\comm{\mathbf{A}}{\mathbf{B}}}$, depending on the parity of the occurrencies of $\mathbf{A}$:
\begin{subequations}
\label{eq:Lemma_principal}
\begin{align}
    &\comm{\left(\mathbf{A}\right)^{2 n +1}}{\mathbf{B}}
    =
    K(\mathbf{A})^{n} \comm{\mathbf{A}}{\mathbf{B}},
    \quad
    n \geq 0
    \\
    &\comm{\left(\mathbf{A}\right)^{2 n +2}}{\mathbf{B}}
    =
    K(\mathbf{A})^{n} \comm{\mathbf{A}}{\comm{\mathbf{A}}{\mathbf{B}}},
    \quad 
    n \geq 0
\end{align}
\end{subequations}
\begin{proof}
First, we prove that every $(2 n + 1)$-commutator is similar to a diagonal matrix, while every $(2 n + 2)$-commutator is similar to a anti-diagonal matrix, for $n\geq 0$.
Since the commutator $\comm{\mathbf{A}}{\mathbf{B}}$ is diagonalizable, it exists a matrix $\mathbf{V}_C$ such that $\mathbf{D}_1 = \mathbf{V}_C^{-1} \comm{\mathbf{A}}{\mathbf{B}} \mathbf{V}_C$. The same similarity transformation applied to the $2$-commutator gives $\mathbf{V}_C^{-1}  \comm{\left(\mathbf{A}\right)^{2}}{\mathbf{B}} \mathbf{V}_C = \comm{\mathbf{V}_C^{-1} \mathbf{A} \mathbf{V}_C }{\mathbf{D}_1} = \widetilde{\mathbf{D}}_2$, which is certainly antidiagonal in two dimensions. 
Analogously, the $3$-commutators maps to $\mathbf{V}_C^{-1}  \comm{\left(\mathbf{A}\right)^{3}}{\mathbf{B}} \mathbf{V}_C = \comm{\mathbf{V}_C^{-1} \mathbf{A} \mathbf{V}_C }{\widetilde{\mathbf{D}}_2} = \mathbf{D}_3$, which is certainly diagonal in two dimensions.
Proceeding by induction, it readily follows that for $n\geq0$:
\begin{subequations}
    \begin{align}
    &\mathbf{D}_{2n+1} = \mathbf{V}_C^{-1} \comm{\left(\mathbf{A}\right)^{2n+1}}{\mathbf{B}} \mathbf{V}_C
    \\
    &\widetilde{\mathbf{D}}_{2n+2} = \mathbf{V}_C^{-1} \comm{\left(\mathbf{A}\right)^{2n+2}}{\mathbf{B}} \mathbf{V}_C
    \end{align}
\end{subequations}
where odd and even labels corresponds to diagonal and anti-diagonal matrices, respectively.

The second part of the proof is based on the observation that, in two dimensions, the matrix $\mathbf{A}' = V_C^{-1} \mathbf{A} V_C$ has equal elements on its diagonal. It can be generically written as:
\begin{align}
    &\mathbf{A}' = 
    \begin{pmatrix}
    a && b \\
    c && a
    \end{pmatrix}.
\end{align}
Thus, if the commutator $\comm{\mathbf{A}}{\mathbf{B}}$ has eigenvalues $\pm d_1$, the anti-diagonal matrix $\widetilde{\mathbf{D}}_2$ is given by:
\begin{align}
    &\widetilde{\mathbf{D}}_2 =
    \comm{\mathbf{A}'}{\mathbf{D}_1}
    =
    2 d_1 
    \begin{pmatrix}
    0 && -b \\
    c && 0
    \end{pmatrix}.
\end{align}
Iterating this procedure and proceeding by induction, it follows that for $n\geq0$:
\begin{subequations}
    \begin{align}
    &\mathbf{D}_{2n+1} = (4 b c)^n \mathbf{D}_1
    \\
    &\widetilde{\mathbf{D}}_{2n+2} = (4 b c)^n \widetilde{\mathbf{D}}_2
    \end{align}
\end{subequations}
Since $4 b c = K(\mathbf{A}') = K(\mathbf{A})$, this completes the proof.
\end{proof}
\end{lemma}
We can now easily obtain the following fundamental results for two dimensional matrix representations:
\begin{lemma}[Hadamard's Lemma]
Let $\mathbf{A}$ and $\mathbf{B}$ be $2\times2$ matrices. Then, the Hadamard's Lemma reads as:
\begin{align}
    \label{eq:HadamardLemma}
    e^{\mathbf{A}} \mathbf{B} 
    e^{-\mathbf{A}} = 
    \mathbf{B} + 
    \frac{\sinh{\left(\Delta_A \right)}}{\Delta_A} \comm{\mathbf{A}}{\mathbf{B}}
    +
    \frac{\cosh{\Delta_A} - 1}{\Delta_A^2}
    \comm{\mathbf{A}}{\comm{\mathbf{A}}{\mathbf{B}}},
\end{align}
where $\Delta_A = \sqrt{K(\mathbf{A})}$ is the difference of the eigenvalues of $A$ and $K(\mathbf{A})$ is the invariant introduced in Lemma~\ref{lemma:principal}.
As a corollary, it follows that:
\begin{align}
    \label{eq:HadamardLemma_exp}
    e^{\mathbf{A}} e^{\mathbf{B}} 
    = 
    e^{\mathbf{B} + 
    \frac{\sinh{\left(\Delta_A \right)}}{\Delta_A} \comm{\mathbf{A}}{\mathbf{B}}
    +
    \frac{\cosh{\Delta_A} - 1}{\Delta_A^2}
    \comm{\mathbf{A}}{\comm{\mathbf{A}}{\mathbf{B}}}}
    e^{\mathbf{A}},
\end{align}
\end{lemma}
\begin{lemma}
\label{lemma:zass}[Zassenhaus formula]
Let $\mathbf{A}$ and $\mathbf{B}$ be $2\times2$ matrices. Then, the exponential of their sum is:
\begin{align}
    \label{eq:Zassenhaus}
    &e^{t\left(\mathbf{A} + \mathbf{B} \right)}
    =
    e^{t \mathbf{A}}
    e^{t \mathbf{B}}
    \,\mathcal{T}
    \text{exp}\Bigg(
    \int_0^t
    d s\,
    \Big\{
    -\frac{\sinh{\left(s \Delta_A \right)}}{\Delta_A} 
    \left(
    \comm{\mathbf{A}}{\mathbf{B}}
    + \frac{\sinh{\left(s \Delta_B \right)}}{\Delta_B}  \comm{\mathbf{A}}{\left(\mathbf{B}\right)^2}
    - \frac{\cosh{\left(s \Delta_B\right)} - 1}{\Delta_B^2}  \comm{\mathbf{A}}{(\mathbf{B})^3}
    \right)
    \notag \\
    &\qquad+
    \frac{\cosh{\left(s \Delta_A\right)} - 1}{\Delta_A^2} \left(
    \comm{\left(\mathbf{A}\right)^2}{\mathbf{B}}
    + \frac{\sinh{\left(s \Delta_B \right)}}{\Delta_B}  \comm{\left(\mathbf{A}\right)^2}{\left(\mathbf{B}\right)^2}
    - \frac{\cosh{\left(s \Delta_B\right)} - 1}{\Delta_B^2}  \comm{\left(\mathbf{A}\right)^2}{\left(\mathbf{B}\right)^3}
    \right)
    \Big\}
    \Bigg)
\end{align}
where $\mathcal{T}$ represents the Dyson time-ordering~\cite{Dyson:1949bp} and $\Delta_A$ and $\Delta_B$ are the differences of the eigenvalues of the matrices $\mathbf{A}$ and $\mathbf{B}$, respectively.
\begin{proof}
The proof follows from the integral representation of the Zassenhaus formula~\cite{Suzuki:1985wzj} and from the iterated application of Lemma~\ref{lemma:principal}. 
\end{proof}
\end{lemma}

%
\bibliographystyle{apsrev4-2}
\bibliography{bibliography}
%

\end{document}